# Field-Temperature Evolution of Antiferromagnetic Phases in Ludvigites Ni$_{3-x}$Mn$_x$BO$_5$

L. N. Bezmaternykh, E. M. Kolesnikova*, E. V. Eremin, S. N. Sofronova, N. V. Volkov, and M. S. Molokeev

Kirensky Institute of Physics, Russian Academy of Sciences, Siberian Branch
Akademgorodok 50, bld. 38, Krasnoyarsk, 660036 Russia
*e-mail: ekoles@iph.krasn.ru

Abstract

The conditions for the flux growth of new Mn-Ni oxyborates with the ludwigite structure are reported. Magnetic measurement data for the samples with nickel and manganese predominance are presented. Diamagnetic anomalies of the antiferromagnetic phases are established and analyzed in the framework of a model comprising two antiferromagnetically interacting subsystems, each being antiferromagnetically ordered.

Introduction

Nowadays, study of magnetism and magnetoelectricity of quasi-low-dimensional structures with partial cation disordering is a topical research direction. Of special interest are transition-metal oxyborates with the ludwigite structures, in which at least one of the transition metals is presented by cations of different valence.

It was shown that the magnetic properties of the ludwigites $Co_2^{2+}Co_{1-x}^{3+}Fe_x^{3+}BO_5$ are highly sensitive to even minor variations in the value of $x$ in the range $0 \leq x \leq 0.15$. Their Neel temperature ($T_N$) can double with increasing $x$. At low temperatures (T = 2–3 K), the coercivity ($H_c$) attains 100 kOe [1–4].

Features of the magnetic behavior of the ludwigites $Co_{2-x}^{2+}Mn_x^{2+}Co_{1-y}^{3+}Mn_y^{3+}BO_5$ also attract much attention [5, 6]. Along with the formation of the antiferromagnetic phase, one can observe traces of a spin-glass-like subsystem even at low concentrations of Mn$^{2+}$ and Mn$^{3+}$ cations. As the Mn concentration is increased, the spin-glass-like phase fraction grows and a magnetic phase transition identified as the transition from the paramagnetic phase to the spin-glass-like state occurs. However, with a further increase in the Mn concentration, the long-range magnetic order recovers. In the antiferromagnetic phase, diamagnetic anomalies of susceptibility were observed upon sample heating after zero magnetic field precooling (ZFC) [5, 6].

These results are interesting for both synthesis and studying the magnetic behavior of the compounds containing cations with other spin-orbit characteristics. In addition, they may be helpful for solving the fundamental problem on the interrelation of the observed macroscopic effects and evolution of local interactions upon variation in the valence state of cations.

The aim of this study was to investigate magnetism of the new ludwigites $Mn_{3-x}Ni_xBO_5$ ($0 < x < 3$). The unique compound of this family with $x = 2.5$ was synthesized by Bluhm et.al. [7], but its magnetic properties have not been studied. Here, we report the magnetic measurement data for the samples with the low ($x_1 = 0.5$) and high ($x_2 = 1.8$) Ni concentration. We compare temperature and field evolutions of the antiferromagnetic phases in these samples and discuss the possibility of their description within one model comprising two antiferromagnetically interacting subsystems, each being antiferromagnetically ordered.

Crystal growth

$Mn_{2.5}Ni_{0.5}BO_5$ ($x_1 = 2.5$ and $n_1 = 15\%$) and $Mn_{1.2}Ni_{1.8}BO_5$ ($x_2 = 1.2$ and $n_2 = 7\%$) single crystals were synthesized from the fluxes

$$(100-n)\% \ mass. \left(Bi_2Mo_3O_{12} + 0.6 \ B_2O_3 + 0.7 \ Na_2O\right) +$$
$$+ n \% \ mass. \left(\frac{(3-x)}{2} Ni_2O_3 + \frac{x}{2} Mn_2O_3 + 0.5 B_2O_3\right)$$

The fluxes in a mass of 50–80 g were prepared from initial trioxides $Mn_2O_3$ and $Ni_2O_3$ in combination with sodium carbonate at the temperature T = 1100°C in a platinum crucible with the volume V = 100 cm³ by sequential melting of powder mixtures, first $Bi_2Mo_3O_{12}$ and $B_2O_3$, then $Mn_2O_3$ and $Ni_2O_3$; finally, $Na_2CO_3$ was added in portions.

In the prepared fluxes, the phase crystallizing within a sufficiently wide (about 40°C) high-temperature range was $Mn_{3-x}Ni_xBO_5$ with the ludwigite structure. The saturation temperatures of the fluxes were $T_{sat1}$ = 920°C and $T_{sat2}$ = 960°C.

Single crystals of the ludwigites were synthesized by spontaneous nucleation. After homogenization of the fluxes at T = 1100°C for 3 h, the temperature was first rapidly reduced to ($T_{sat}$–10)°C and then slowly reduced with a rate of 2–4°C/day. In 3 days, the growth was completed, the crucible was withdrawn from the furnace, and the flux was poured out. The grown single crystals in the form of orthogonal prisms with a length of 10 mm and a transverse size of about 0.5 mm were etched in a 20% water solution of nitric acid to remove the flux remainder.

Structural data

X-ray investigations of the $Mn_{2.5}Ni_{0.5}BO_5$ single crystal and $Mn_{1.2}Ni_{1.8}BO_5$ powder were carried out on a SMART APEXII diffractometer (Mo $K_\alpha$, $\lambda = 0.7106$ Å) at room temperature. The obtained data are given in Table 1. Both $Mn_{2.5}Ni_{0.5}BO_5$ and $Mn_{1.2}Ni_{1.8}BO_5$ samples belong to the space group Pbam ($D_{2h}^9$), i.e., have the ludwigite structure. The structure was refined by the least-squares minimization using SHELX97 [8]. The unit cell involves four formula units, i.e., contains 12 magnetic atoms occupying 4 nonequivalent positions: 4g, 4h, 2a, and 2d. The ludwigite structure is presented in Fig. 1. We investigated occupation of the crystallographic positions in $Mn_{2.5}Ni_{0.5}BO_5$ by magnetic atoms and showed that the positions 4h, 2a, and 2d are occupied only by Mn. Nickel and manganese ions together occupy the 4g position in $Mn_{2.5}Ni_{0.5}BO_5$, however, we failed to calculate the occupancies of this position by each ion because of the similarity of their atomic functions (Table 2). In $Mn_{1.2}Ni_{1.8}BO_5$, all the crystallographic positions are occupied by Ni and Mn ions; their refined occupancies are given in Table 3.

Magnetic characterization

Magnetic properties of the $Mn_{2.5}Ni_{0.5}BO_5$ single crystal and the $Mn_{1.2}Ni_{1.8}BO_5$ sample consisting of several c-axes-oriented crystals were measured on a PPMS-9 Physical Property Measurement System (Quantum Design) at the temperatures T = 3–300 K in the magnetic fields H = 0.1–80 kOe.

Temperature dependences of magnetization for the investigated samples are presented in Figs. 2 and 3. The dependences were obtained upon cooling the sample in the magnetic field H = 1 kOe (FC) parallel (H ∥ c) or orthogonal (H ⊥ c) to the c axis. The magnetization of the $Mn_{2.5}Ni_{0.5}BO_5$ crystal (Fig. 2) monotonically increases below $T_N = 81$ K at H ⊥ c. In case H ∥ c, the temperature range of the magnetization variation is much narrower. Presumably, near $T_N = 81$ K the phase transition from the paramagnetic to antiferromagnetic state occurs, which is related to alignment of the magnetic moments in the planes perpendicular to the c axis. The slow magnetization growth near $T_N$ can result from the almost collinear alignment of the magnetic moments in this region. The temperature dependence of the inverse susceptibility $\chi_\perp^{-1} = \dfrac{H_\perp}{M}$ illustrated in the inset in Fig. 2 also indicates predominance of the antiferromagnetic interaction. According to this dependence, the paramagnetic Curie temperature is negative: $\theta = -40$ K. In the paramagnetic phase, no magnetic anisotropy was found.

The paramagnetic Curie temperature of the $Mn_{1.2}Ni_{1.8}BO_5$ sample (Fig. 3) is also negative. The antiferromagnetic ordering temperature $T_N = 92$ K was estimated from the temperature dependence of the magnetization obtained at $H \perp c$. The rapid growth of M near $T_N$ can be attributed to the noncollinear alignment of the magnetic moments. Minor variations in M at $H \parallel c$ indicate that the spin-lattice interaction fixes the magnetic moments in the planes perpendicular to the c axis. The magnetic moments tend to orient in this way already in the paramagnetic phase.

Another important feature of the temperature dependence of the magnetization for the $Mn_{1.2}Ni_{1.8}BO_5$ sample at $H \perp c$ is the presence of the compensation point $M = 0$. Below this point, the susceptibility $\chi_\perp^{-1} = \dfrac{H_\perp}{M}$ is negative. As the temperature is decreased, the absolute value of the susceptibility increases. Such an anomaly of the susceptibility in the antiferromagnetic phase can be considered as diamagnetic.

In the other regime (ZFC), temperature dependences of the magnetization were obtained upon sample heating after zero field cooling. The ZFC and FC dependences obtained at $H \perp c$ for two samples strongly differ (Figs. 4a and 4b).

In the ZFC regime, the $Mn_{2.5}Ni_{0.5}BO_5$ sample passes to the state with the negative susceptibility at switching on the magnetic field (Fig. 4a). As the temperature is increased, the absolute value of the susceptibility decreases. At certain temperature $T_{cr}$, the magnetic moment reverses and the sample undergoes a transition to the state obtained in the FC regime. With increasing magnetic field H, $T_{cr}$ decreases (Fig. 5). In strong magnetic fields, the ZFC and FC dependences coincide. These measurements revealed the existence of two possible states of the antiferromagnetic phase in the $Mn_{2.5}Ni_{0.5}BO_5$ sample. The bistability is confirmed by the magnetic field dependences of the magnetization at different temperatures (Fig. 6). The temperature dependence of coercivity $H_c$ is consistent with the field dependence of critical temperature $T_{cr}$ (Fig. 7). Therefore, the key role in the formation of these two states is played by the spin-lattice interaction.

It should be noted that at $H \parallel c$ the magnetic hysteresis is not observed (Fig. 8).

The magnetic behavior of $Mn_{2.5}Ni_{0.5}BO_5$ described within the model of two antiferromagnetically interacting subsystems can be explained as follows. Subsystem $M_I$ is characterized by the stronger spin-lattice coupling, i.e., the higher coercivity, but its magnetic-field-induced magnetization is lower than that in subsystem $M_{II}$. Subsystem $M_{II}$ is

"softer". Due to the strong antiferromagnetic interaction of these subsystems in weak fields, the sample can be in one of the two possible states: with the resulting induced moment $M = M_{II} - M_I$ directed either along the magnetic field vector or oppositely. In the FC regime, below $T_N$ the state is implemented where the magnetic moment of subsystem $M_{II}$ is directed along the magnetic field; correspondingly, the smaller magnetic moment of subsystem $M_I$ is directed oppositely. The coercivity of subsystem $M_I$ at these temperatures is small. As the temperature is decreased, the difference between these moments monotonically increases. The coercivity determined mainly by subsystem $M_I$ also increases. In the ZFC regime, the moment of subsystem $M_I$, due to the shorter relaxation time, aligns along the magnetic field and specifies the growth of the moment of subsystem $M_{II}$ in the opposite direction. This state with the resulting magnetic moment directed oppositely to the magnetic field is stable only at $T < T_{cr}$.

The $Mn_{1.2}Ni_{1.8}BO_5$ sample (Fig. 4b) in the ZFC regime undergoes a transition to the state with the positive susceptibility at switching on the magnetic field. With an increase in temperature, the magnetization decreases, passes the compensation point $M = 0$, and near $T_N$ turns to the FC state. The alternating behavior of the susceptibility is observed only in weak magnetic fields (Fig. 9). Magnetic field dependences of the magnetization for this sample differ from those for $Mn_{2.5}Ni_{0.5}BO_5$ (Figs. 10 and 11). At $H \perp c$ and low temperatures, the hysteresis loops are strongly extended and the difference between their branches vanishes only in strong magnetic fields ($H = 60-70$ kOe). As the temperature is increased, the loop shape changes and at $T = 50$ K the loop is similar to that observed for $Mn_{2.5}Ni_{0.5}BO_5$. At $H \parallel c$, there is no hysteresis (Fig. 11).

In terms of crystal chemistry, the established features of the magnetic hysteresis of the $Mn_{1.2}Ni_{1.8}BO_5$ sample can be attributed to the higher degree of positional disordering of $Ni^{2+}$ cations as compared with the $Mn_{2.5}Ni_{0.5}BO_5$ sample. These cations with the strong spin-orbit coupling determine the spin-lattice interaction of the magnetic subsystems. Disordering of $Ni^{2+}$ cations is accompanied by the formation of fragments with mutually misoriented easy axes in the antiferromagnetic phase. This yields the extended hysteresis loop in the low-temperature region. However, with an increase in temperature, the spin-lattice coupling weakens and the axis characteristic of the sample with the low Ni concentration becomes predominant.

The model comprising two antiferromagnetically interacting subsystems each being antiferromagnetically ordered can be adapted to the $Mn_{1.2}Ni_{1.8}BO_5$ crystal. In this case, it is assumed that each antiferromagnetic subsystem in the model reveals the properties described in the previous section, the coercivities of the subsystems are comparable and the sign of the magnetization $M = M_I - M_{II}$ changes with temperature. In the FC regime, in the temperature region from Neel temperature $T_N$ to compensation point $T_{cr}$ the resulting magnetization is $M = M_I - M_{II} > 0$ ($\chi > 0$); below the compensation point, it is $M = M_I - M_{II} < 0$ ($\chi < 0$). In the ZFC regime, after switching on the magnetic field the state with $M = M_I - M_{II} > 0$ ($\chi > 0$) is stabilized. As the temperature is increased, the resulting magnetization below compensation point $T_{cr}$ is $M = M_I - M_{II} < 0$ ($\chi < 0$). After that, at $T = T_{cr}$, the spin rotates by 180° and the crystal passes to the FC state (inset in Fig. 4). This transition is similar to the transition observed in the $Mn_{2.5}Ni_{0.5}BO_5$ sample in the ZFC regime at H ⊥ c. In both cases, spin reorientation occurs in the plane perpendicular to the c axis and is accompanied by the change in the susceptibility sign. The alternating behavior of the susceptibility of the antiferromagnetic phase in the $Mn_{1.2}Ni_{1.8}BO_5$ sample is analogous to that observed on a powder sample of the Ni-containing antiferromagnet $Ni(HCOO)_2 \cdot 2H_2O$ [9].

Conclusions

The existence of the ludwigites $Mn_{3-x}Ni_xBO_5$ ($0 < x < 3$) with heterovalent Mn and Ni cations was established. Two qualitatively different scenarios in the behavior of the antiferromagnetic phases for the samples with the low ($x_1 = 0.5$) and high ($x_2 = 1.8$) Ni concentrations with the diamagnetic anomalies of the susceptibility were revealed. The possibility of describing these scenarios in the framework of the model comprising two antiferromagnetically interacting subsystems, each being antiferromagnetically ordered, was demonstrated.

To understand the correlation between these anomalies and the cation concentration in the ludwigite crystals, it is important to study the features of the magnetic behavior of the ludwigites $Mn_{3-x}Ni_xBO_5$ with $0.5 < x < 1.8$ and $x > 3$.

Acknowledgements
The study was supported by the Siberian Branch of the Russian Academy of Sciences, integration project no. 29 and project no. 2.5.2.

References
[1] J. Bartolomé, A. Arauzo et.al. Phys. Rev. B. – 2011. – V. 83. –P. 144426.
[2] Yu.V. Knyazev, N.B. Ivanova et.al. JMMM. – 2012. – V. 324. – P. 923-927.
[3] D.C. Freitas, M.A. Continentino et.al. Phys. Rev. B. – 2008. – V. 77. – P. 184422.

[4] M. S. Platunov, S. G. Ovchinnikov et. al. JETP Letters. – 2012. – V. 96 (10). – P. 650-654.

[5] L.N. Bezmaternykh, E.M. Kolesnikova et.al. Abstract volume of III International conference Crystallogenesis and Mineralogy. – 2013. – P. 67-68.

[6] Yu.V. Knyazev, N.B. Ivanova et.al. JMMM. – 2012. – V. 324. – P. 923-927.

[7] K. Bluhm, Hk. Müller-Buschbaum. Z. anorg. allg. Chem. – 1989. – V. 579. – P. 111-115.

[8] G. M. Sheldrick. Acta Crystallographica Section A. – 1990. – V. 46. – P. 467–473.

[9] H. Kageyama, D.I. Khomskii et.al. Phys. Rev. B. – 2003. – V. 67. – P. 224422.

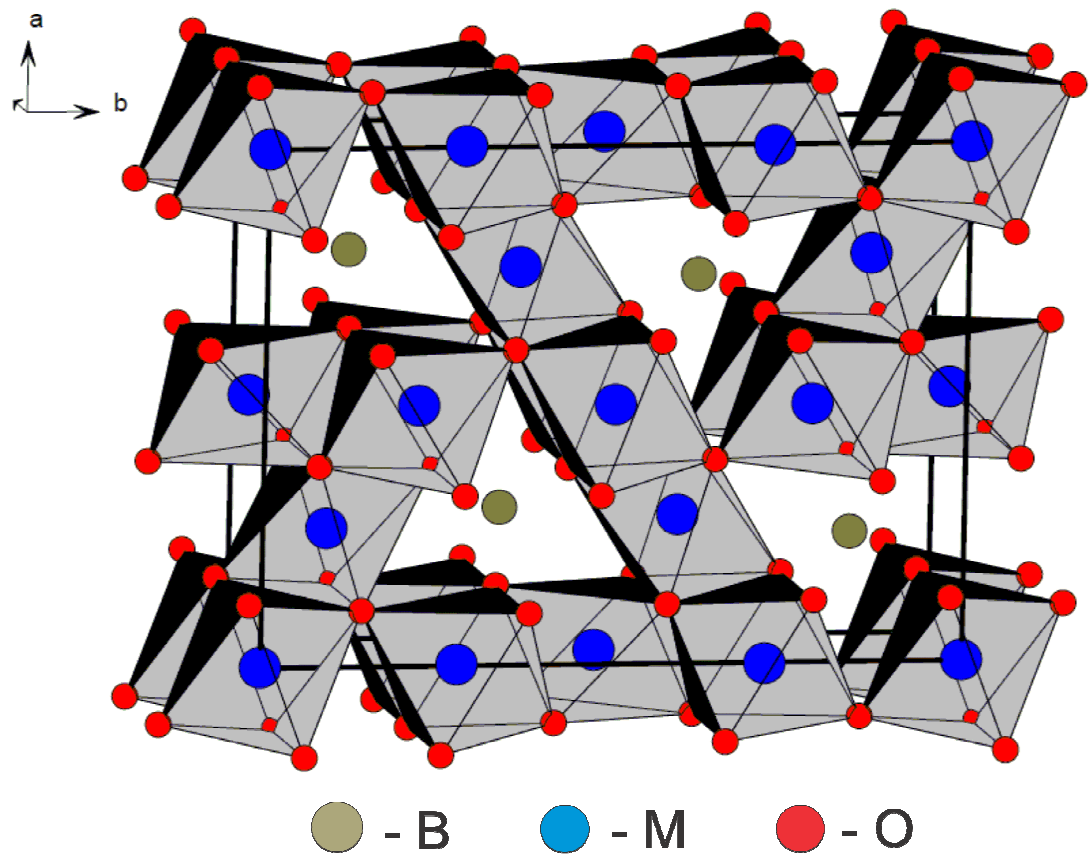

Fig. 1 Ludwigite structure (B – boron; M – transition metal (Mn or Ni); O – oxygen).

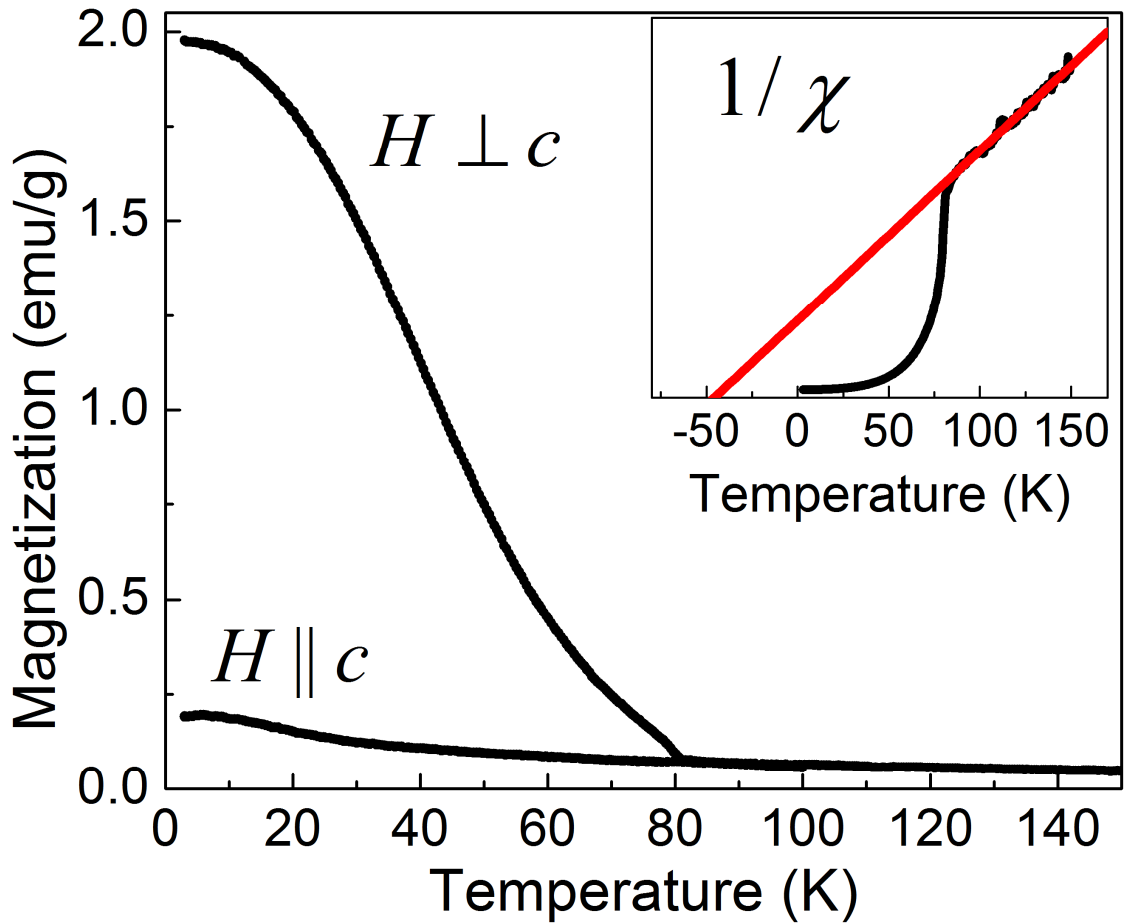

Fig. 2 Temperature dependencies of magnetization of Mn$_{2.5}$Ni$_{0.5}$BO$_5$ at different orientations of magnetic field ($H=1$ kOe; $H \perp c$, $H // c$). Inset: temperature dependency of inverse susceptibility $\chi_\perp^{-1}$ (black line) and its linear extrapolation of paramagnetic area (red line).

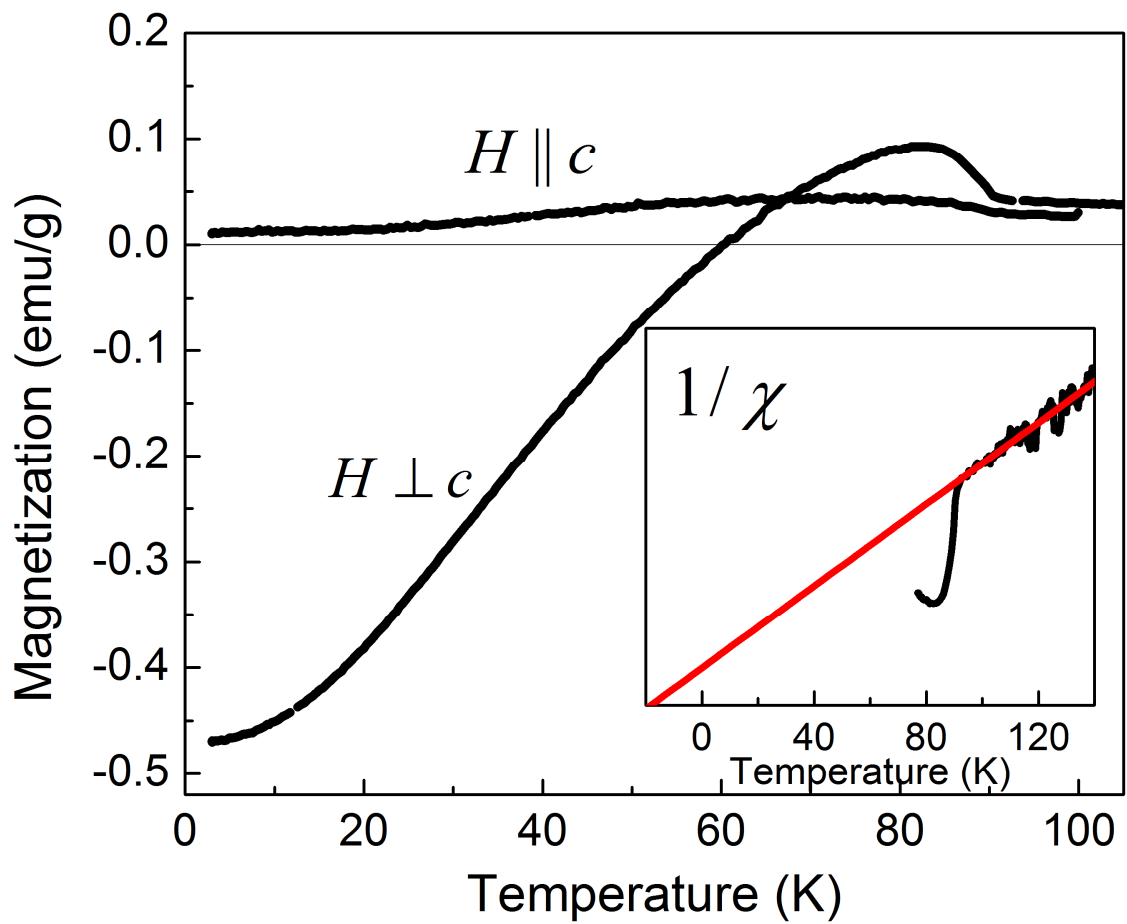

Fig. 3 Temperature dependencies of magnetization of $Mn_{1.2}Ni_{1.8}BO_5$ at different orientations of magnetic field ($H=1\ kOe$; $H \perp c$, $H // c$). Inset: temperature dependency of inverse susceptibility $\chi_\perp^{-1}$ (black line) and its linear extrapolation of paramagnetic phase (red line).

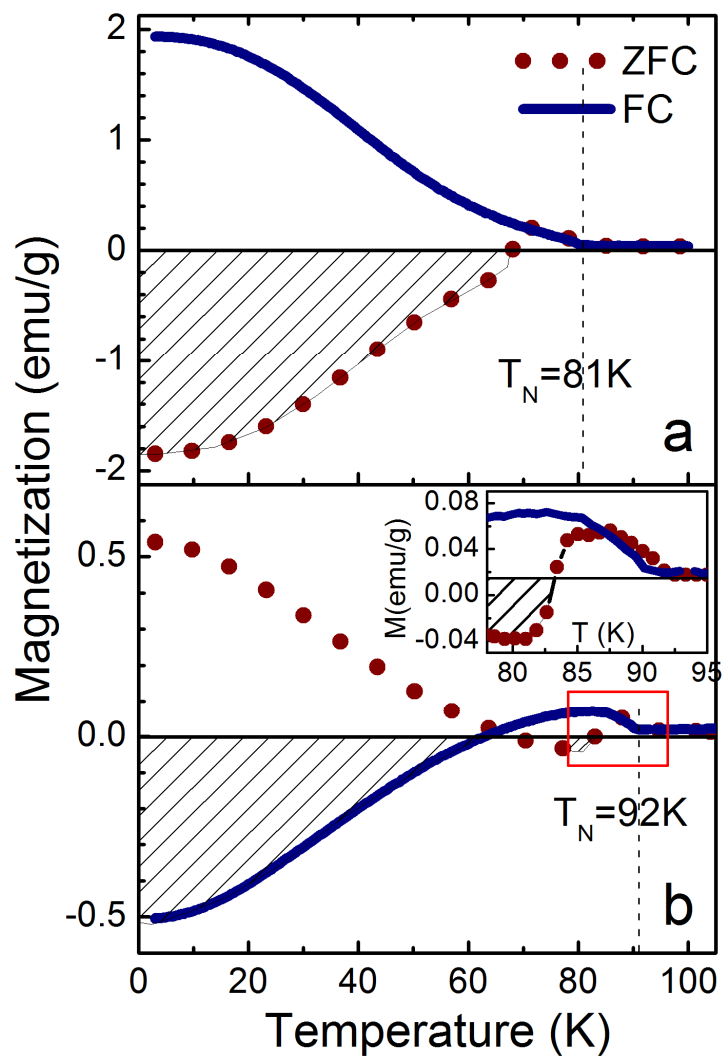

Fig. 4 Temperature FC- and ZFC-dependencies of magnetization measured at *H=0.5 kOe*, $H \perp c$ (a – $Mn_{2.5}Ni_{0.5}BO_5$; b – $Mn_{1.2}Ni_{1.8}BO_5$).

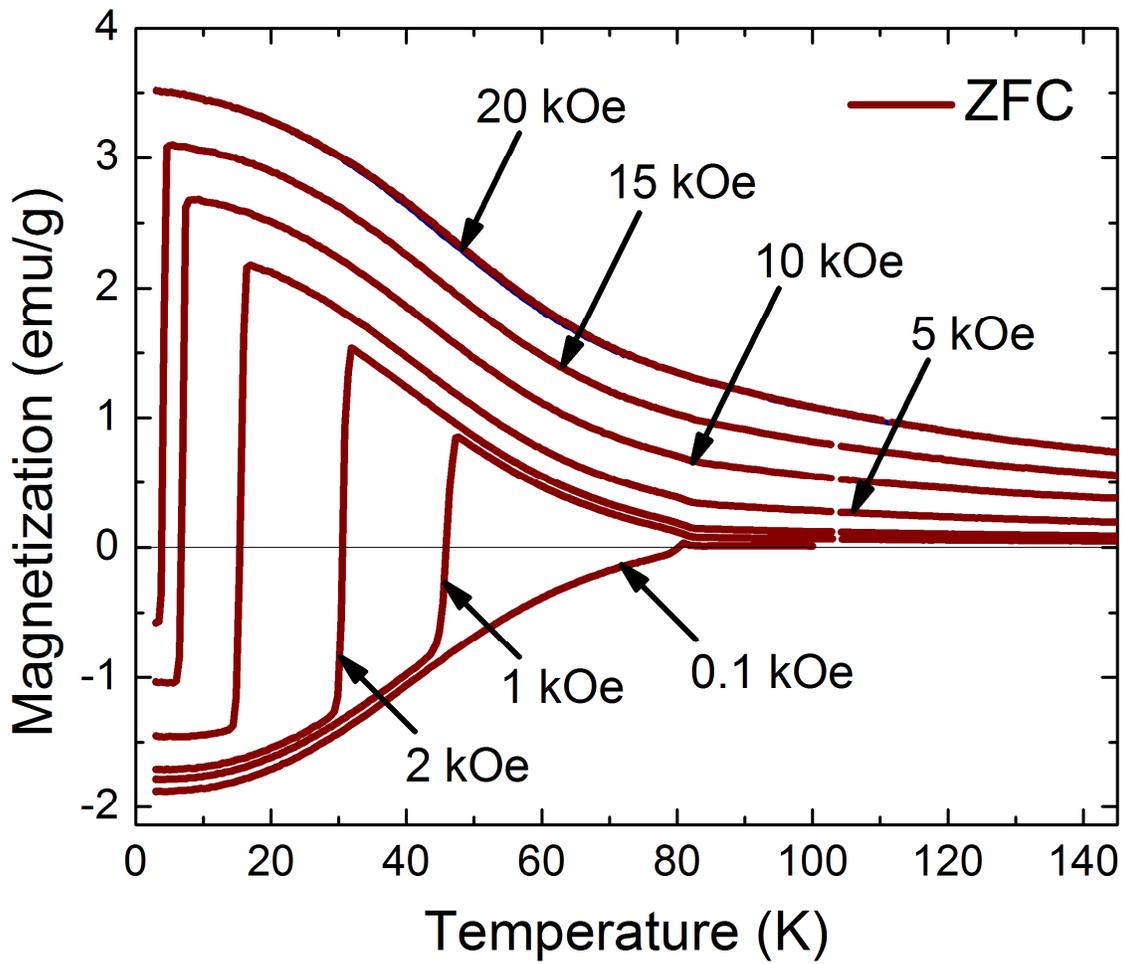

Fig. 5 Temperature ZFC-dependencies of magnetization of $Mn_{2.5}Ni_{0.5}BO_5$ measured at different magnetic fields, $H \perp c$.

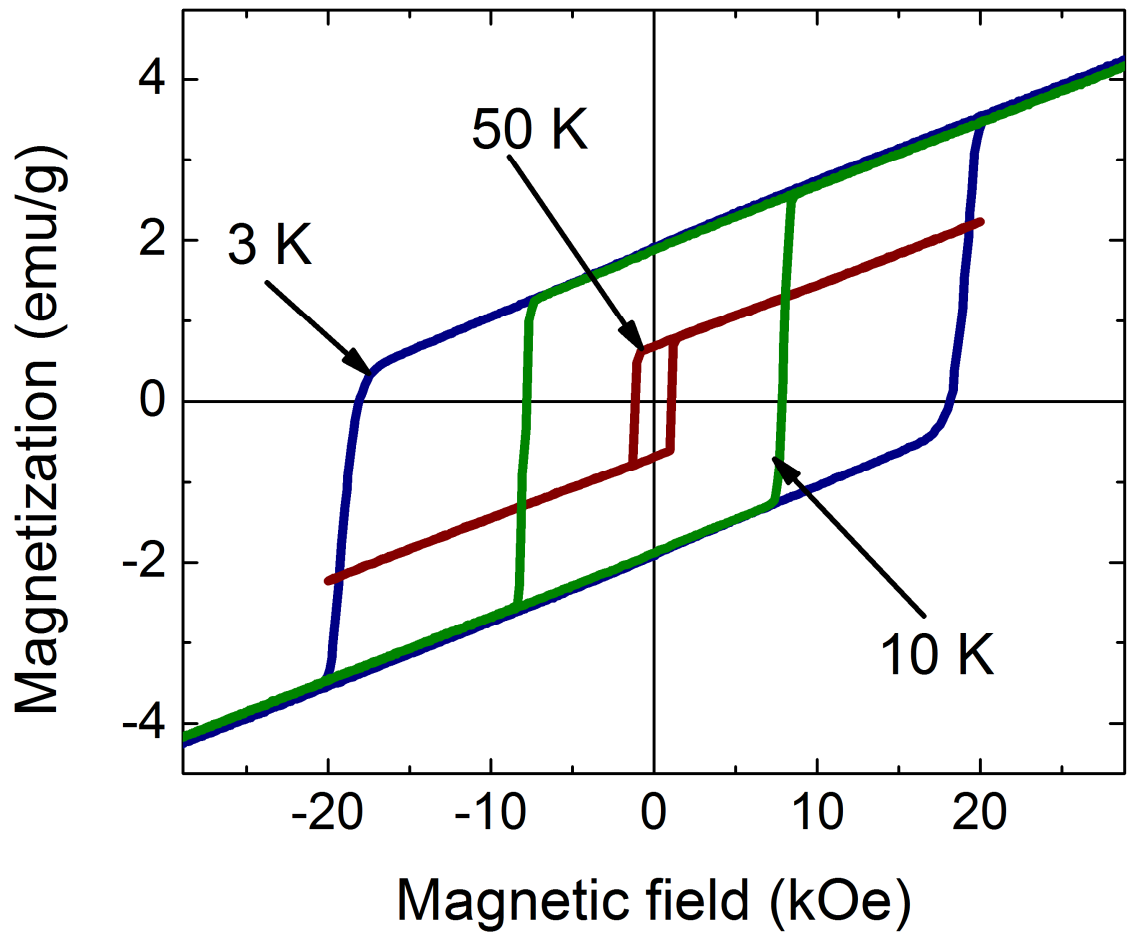

Fig. 6 Magnetic field dependencies of magnetization of $Mn_{2.5}Ni_{0.5}BO_5$ measured at different temperatures ($H \perp c$).

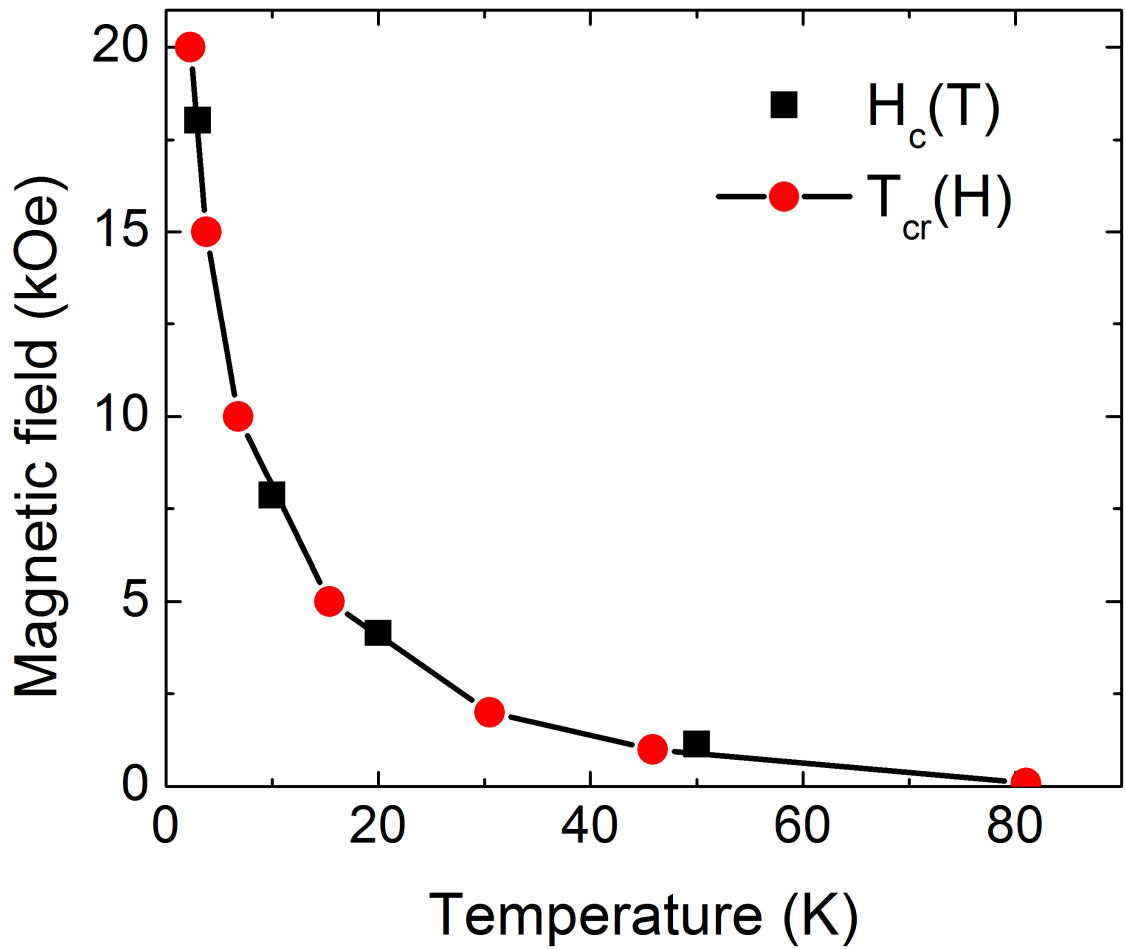

Fig. 7 Temperature dependency of coercitive field $H_c$ (black squares) and dependency of critical temperature $T_{cr}$ (red circles) of magnetic field of $Mn_{2.5}Ni_{0.5}BO_5$.

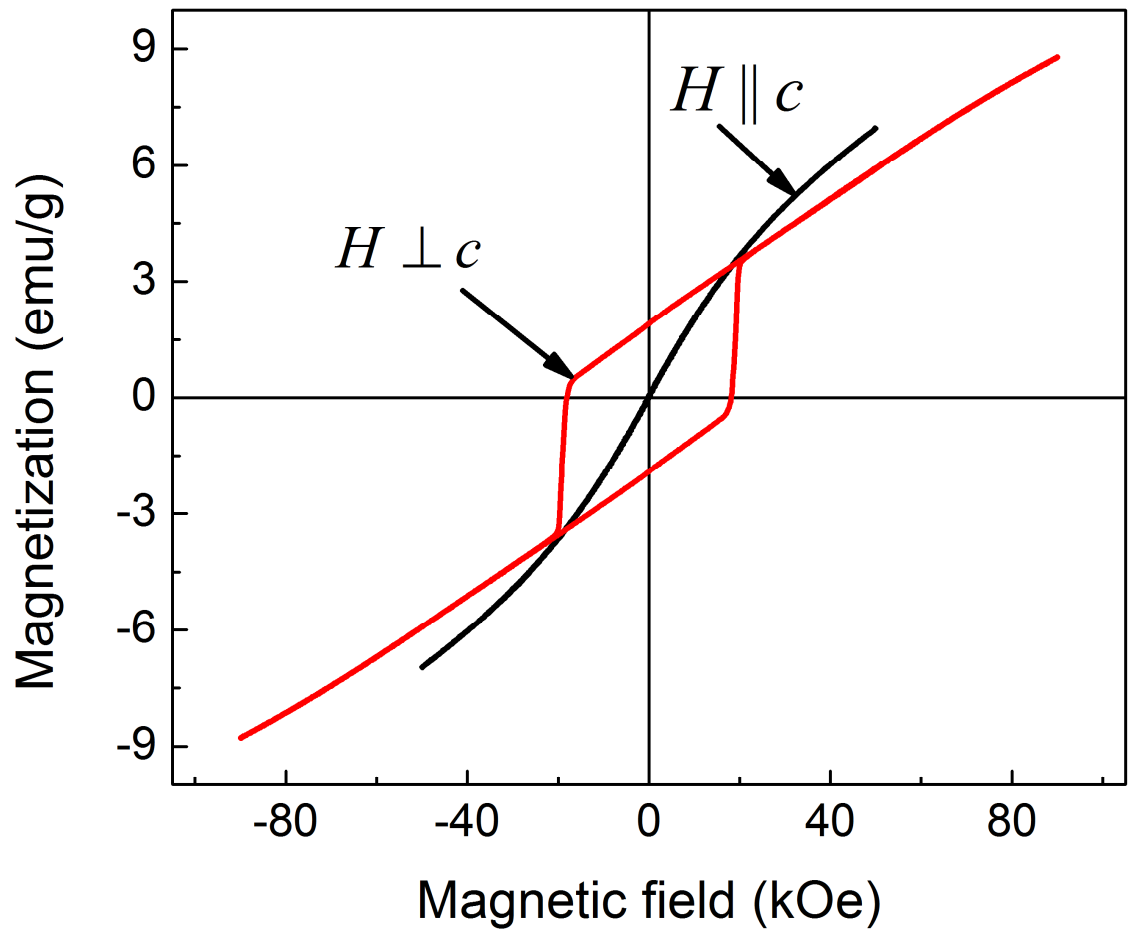

Fig. 8 Magnetic field dependencies of magnetization of $Mn_{2.5}Ni_{0.5}BO_5$ measured at $H \perp c$ and $H // c$ (T=3 K).

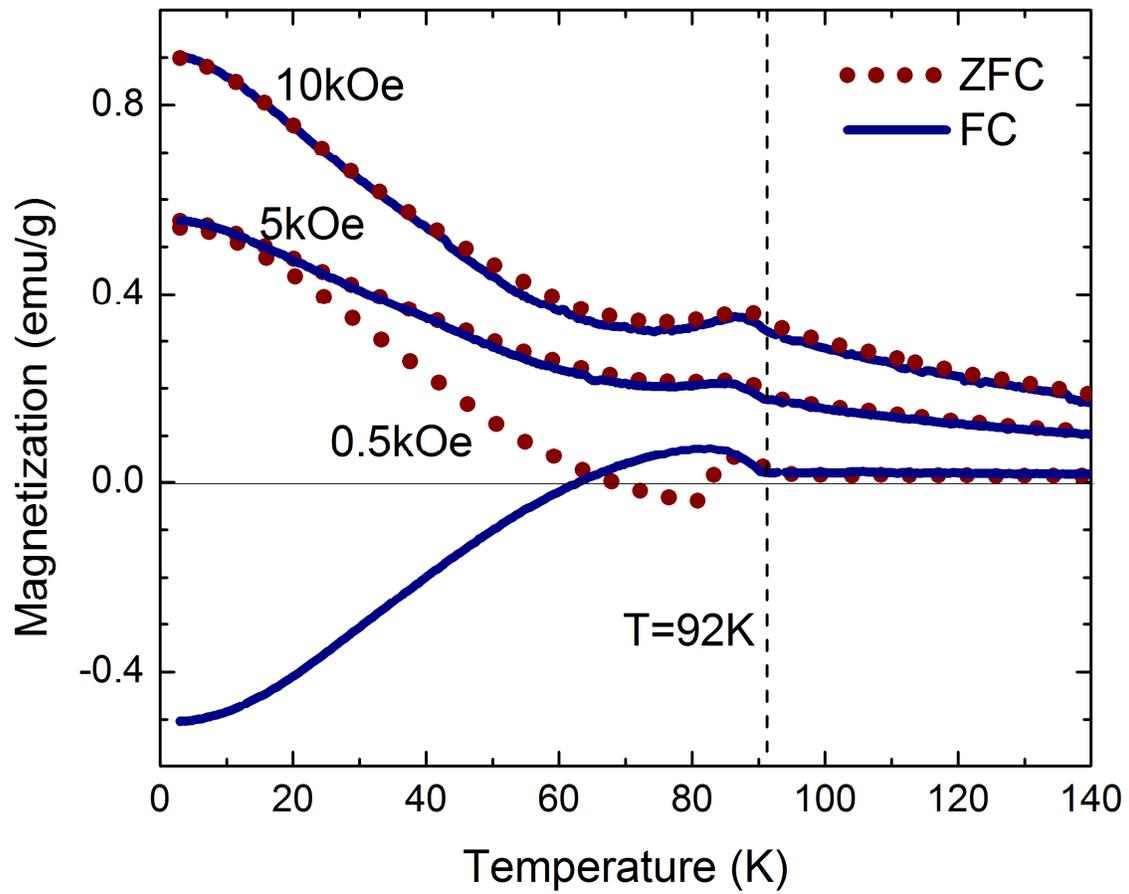

Fig. 9 Temperature FC- and ZFC-dependencies of magnetization of $Mn_{1.2}Ni_{1.8}BO_5$ measured at different magnetic fields ($H \perp c$).

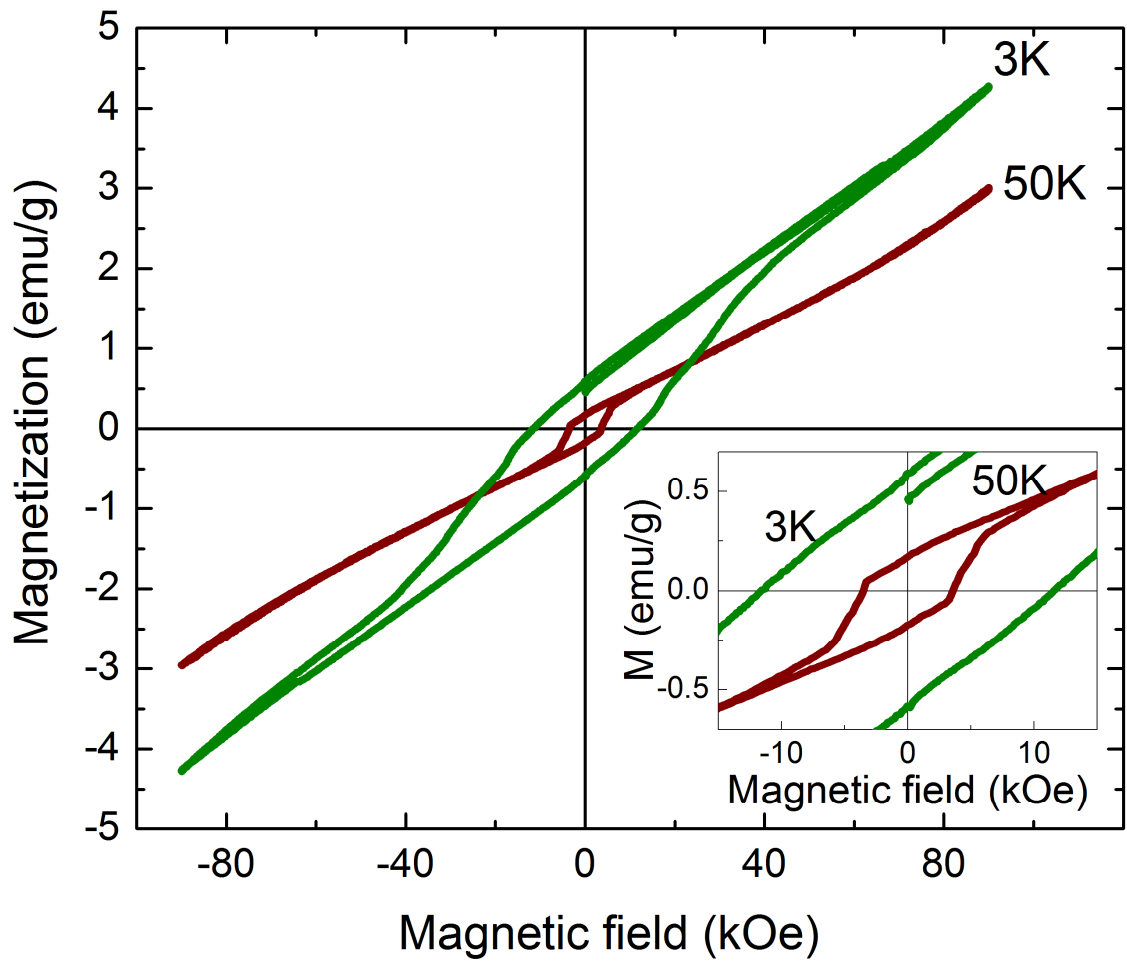

Fig. 10 Magnetic field dependencies of magnetization of Mn$_{1.2}$Ni$_{1.8}$BO$_5$ measured at different temperatures ($H \perp c$).

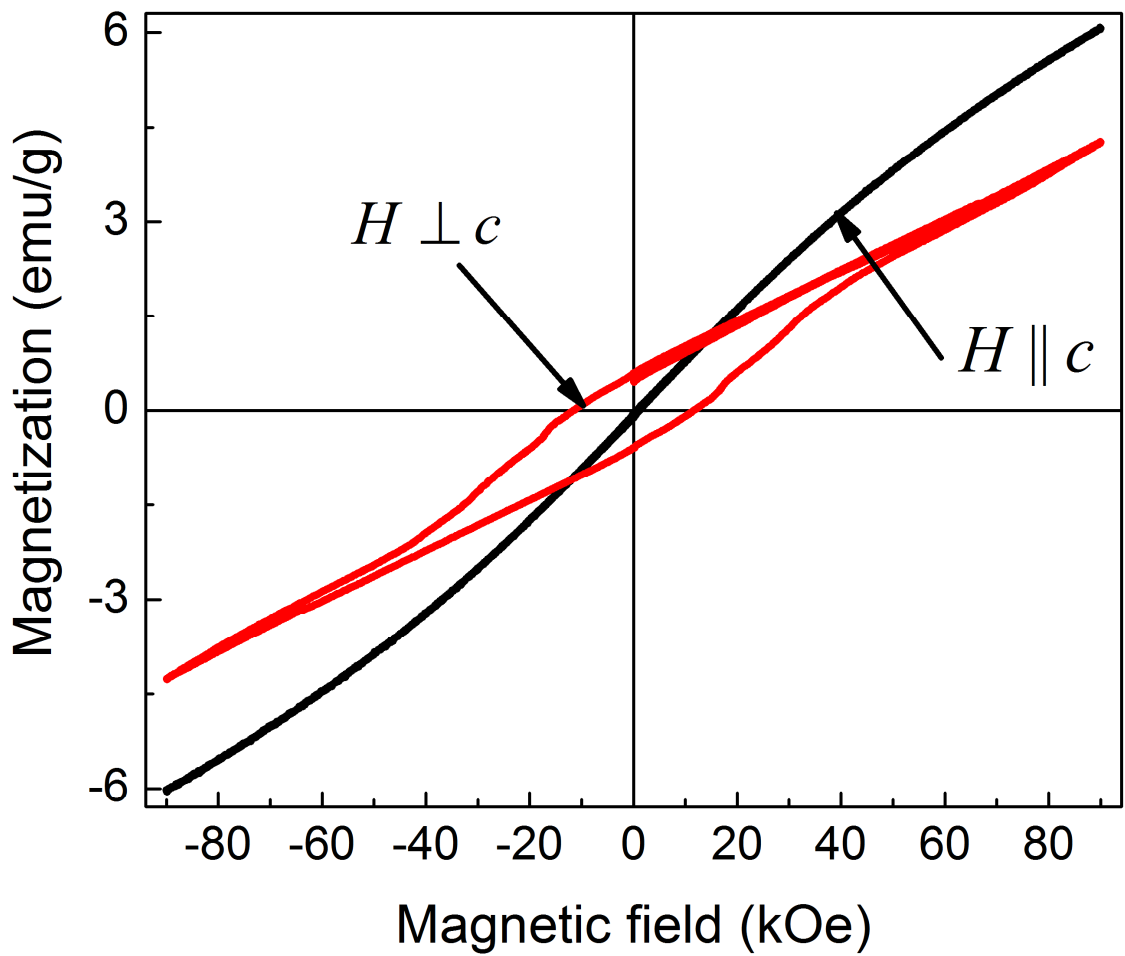

Fig. 11 Magnetic field dependencies of magnetization of $Mn_{1.2}Ni_{1.8}BO_5$ measured at $H \perp c$ and $H // c$ (T=3 K).

Table 1. Crystallographic data and main processing and refinement parameters for
$Mn_{2.5}Ni_{0.5}BO_5$ and $Mn_{1.2}Ni_{1.8}BO_5$

| Crystallographic data | | |
|---|---|---|
| Chemical formula | $Mn_{2.5}Ni_{0.5}BO_5$ | $Mn_{1.2(1)}Ni_{1.8(1)}BO_5$ |
| $M_r$ | 257.51 | 261.49 |
| Space group, Z | *Pbam*, 4 | *Pbam*, 4 |
| $a$, (Å) | 9.179(2) | 9.187(1) |
| $b$, (Å) | 12.344(2) | 12.322(1) |
| $c$, (Å) | 3.0010(6) | 3.0010(3) |
| $V$, (Å$^3$) | 340.0(1) | 339.71(6) |
| $D_x$, Mg/m$^3$ | 5.030 | 5.113 |
| $\mu$, mm$^{-1}$ | 11.769 | 13.952 |
| Size | 0.1×0.1×0.5 mm | 0.1×0.1×0.5 mm |
| Data collection | | |
| Wavelength | MoK$_\alpha$, $\lambda = 0.7106$ Å | MoK$_\alpha$, $\lambda = 0.7106$ Å |
| Measured reflections | 3107 | 3092 |
| Independent reflections | 546 | 546 |
| Reflections with I>2σ(I) | 514 | 503 |
| Absorption correction | Multiscan | Multiscan |
| $R_{int}$ | 0.0409 | 0.0406 |
| $2\theta_{max}$ (°) | 59.28 | 59.08 |
| $h$ | -12 → 12 | -12 → 12 |
| $k$ | -17 → 16 | -16 → 16 |
| $l$ | -4 → 4 | -4 → 4 |
| Refinement | | |
| $R[F^2>2\sigma(F^2)]$ | 0.0381 | 0.0306 |
| $wR(F^2)$ | 0.1112 | 0.0711 |
| $S$ | 1.007 | 1.039 |
| Weight | $w=1/[\sigma^2(F_o^2)+(0.00744P)^2+2.63P]$ where $P=\max(F_o^2+2F_c^2)/3$ | $w=1/[\sigma^2(F_o^2)+(0.0457P)^2+102P]$ where $P=\max(F_o^2+2F_c^2)/3$ |
| $(\Delta/\sigma)_{max}$ | <0.07 | <0.01 |
| $\Delta\rho_{max}$, e/Å$^3$ | 1.77 | 1.02 |
| $\Delta\rho_{min}$, e/Å$^3$ | -1.50 | -1.00 |
| Extinction correction coefficient (SHELX97) | 0.098(8) | 0.062(4) |

Table 2. Fractional atomic coordinates and isotropic or equivalent isotropic displacement parameters (Å$^2$) for Mn$_{2.5}$Ni$_{0.5}$BO$_5$

|     | Wyck. | x | y | z | $U_{iso}$*/$U_{eq}$ | Occ. (<1) |
|---|---|---|---|---|---|---|
| Mn1 | 4g | -0.00269 (8) | 0.71967 (7) | 0.0000 | 0.0096 (3) | 0.50 |
| Ni1 | 4g | -0.00269 (8) | 0.71967 (7) | 0.0000 | 0.0096 (3) | 0.50 |
| Mn2 | 2a | 0.0000 | 1.0000 | 0.0000 | 0.0046 (4) | |
| Mn3 | 4h | 0.25977 (9) | 0.61538 (7) | -0.5000 | 0.0069 (3) | |
| Mn4 | 2d | 0.0000 | 0.5000 | -0.5000 | 0.0050 (4) | |
| O1 | 4g | -0.1067 (5) | 0.8564 (3) | 0.0000 | 0.0166 (10) | |
| O2 | 4h | 0.1449 (4) | 0.7642 (4) | -0.5000 | 0.0149 (9) | |
| O3 | 4g | 0.1128 (5) | 0.5796 (4) | 0.0000 | 0.0167 (9) | |
| O4 | 4h | -0.1272 (5) | 0.6416 (3) | -0.5000 | 0.0129 (9) | |
| O5 | 4h | 0.1477 (5) | 0.9582 (3) | -0.5000 | 0.0142 (9) | |
| B | 4h | 0.2225 (8) | 0.8618 (5) | -0.5000 | 0.0116 (13)* | |

Table 3. Fractional atomic coordinates and isotropic or equivalent isotropic displacement parameters (Å$^2$) for Mn$_{1.2(1)}$Ni$_{1.8(1)}$BO$_5$

|     | Wyck. | x | y | z | $U_{iso}$*/$U_{eq}$ | Occ. (<1) |
| --- | --- | --- | --- | --- | --- | --- |
| Ni1 | 4g | -0.00214 (5) | 0.78052 (5) | 0.0000 | 0.0071 (3) | 0.68 (4) |
| Mn1 | 4g | -0.00214 (5) | 0.78052 (5) | 0.0000 | 0.0071 (3) | 0.31 (4) |
| Ni2 | 2a | 0.0000 | 0.5000 | 0.0000 | 0.0071 (3) | 0.69 (4) |
| Mn2 | 2a | 0.0000 | 0.5000 | 0.0000 | 0.0071 (3) | 0.31 (4) |
| Ni3 | 4h | 0.26000 (6) | 0.88485 (5) | -0.5000 | 0.0069 (2) | 0.34 (4) |
| Mn3 | 4h | 0.26000 (6) | 0.88485 (5) | -0.5000 | 0.0069 (2) | 0.66 (4) |
| Ni4 | 2d | 0.0000 | 1.0000 | -0.5000 | 0.0076 (3) | 0.69 (5) |
| Mn4 | 2d | 0.0000 | 1.0000 | -0.5000 | 0.0076 (3) | 0.31 (5) |
| O1 | 4g | -0.1066 (3) | 0.6436 (2) | 0.0000 | 0.0129 (7) | 1 |
| O2 | 4h | 0.1129 (3) | 0.9209 (3) | 0.0000 | 0.0138 (7) | 1 |
| O3 | 4g | 0.1471 (3) | 0.7363 (2) | -0.5000 | 0.0118 (7) | 1 |
| O4 | 4h | -0.1264 (3) | 0.8588 (2) | -0.5000 | 0.0105 (7) | 1 |
| O5 | 4h | 0.1475 (3) | 0.5421 (2) | -0.5000 | 0.0112 (7) | 1 |
| B1 | 4h | -0.2765 (6) | 0.8608 (4) | -0.5000 | 0.0104 (10)* | 1 |